\documentclass[twocolumn]{aastex62}
\usepackage{amssymb}
\usepackage{chngpage}

\shorttitle{Radio Transients from SLSN-I}
\shortauthors{Law et al.}

\begin{document}

\title{A Search for Late-Time Radio Emission and Fast Radio Bursts from Superluminous Supernovae}

\author[0000-0002-4119-9963]{C.~J.~Law}
\affiliation{Department of Astronomy and Owens Valley Radio Observatory, California Institute of Technology, Pasadena, CA 91125, USA}
\affiliation{Department of Astronomy and Radio Astronomy Lab, University of California, Berkeley, CA 94720, USA}

\author[0000-0002-9646-8710]{C.~M.~B.~Omand}
\affiliation{Department of Physics, School of Science, the University of Tokyo, Tokyo 113-0033, Japan}

\author{K.~Kashiyama}
\affiliation{Department of Physics, School of Science, the University of Tokyo, Tokyo 113-0033, Japan}
\affiliation{Research Center for the Early Universe, the University of Tokyo, Tokyo 113-0033, Japan}

\author[0000-0002-5358-5642]{K.~Murase}
\affiliation{Department of Physics, The Pennsylvania State University, University Park, PA 16802, USA}
\affiliation{Department of Astronomy \& Astrophysics, The Pennsylvania State University, University Park, PA 16802, USA}
\affiliation{Center for Particle and Gravitational Astrophysics, The Pennsylvania State University, University Park, PA 16802, USA}
\affiliation{Yukawa Institute for Theoretical Physics, Kyoto University, Kyoto 606-8502, Japan}

\author[0000-0003-4056-9982]{G.~C.~Bower}
\affiliation{Academia Sinica Institute of Astronomy and Astrophysics, 645 N. A'ohoku Place, Hilo, HI 96720, USA}

\author[0000-0002-2059-0525]{K.~Aggarwal}
\affiliation{Department of Physics and Astronomy, West Virginia University, Morgantown, WV 26506, USA}
\affiliation{Center for Gravitational Waves and Cosmology, West Virginia University, Chestnut Ridge Research Building, Morgantown, WV 26505}

\author{S.~Burke-Spolaor}
\affiliation{National Radio Astronomy Observatory, Socorro, NM 87801, USA}
\affiliation{Department of Physics and Astronomy, West Virginia University, Morgantown, WV 26506, USA}
\affiliation{Center for Gravitational Waves and Cosmology, West Virginia University, Chestnut Ridge Research Building, Morgantown, WV 26505}

\author{B.~J.~Butler}
\affiliation{National Radio Astronomy Observatory, Socorro, NM 87801, USA}

\author{P.~Demorest}
\affiliation{National Radio Astronomy Observatory, Socorro, NM 87801, USA}

\author{T.~J.~W.~Lazio}
\affiliation{Jet Propulsion Laboratory, California Institute of Technology, Pasadena, CA 91109, USA}

\author{J.~Linford}
\affiliation{Department of Physics and Astronomy, West Virginia University, Morgantown, WV 26506, USA}
\affiliation{Center for Gravitational Waves and Cosmology, West Virginia University, Chestnut Ridge Research Building, Morgantown, WV 26505}

\author[0000-0003-2548-2926]{S.~P.~Tendulkar}
\affiliation{Department of Physics \& McGill Space Institute, McGill University, 3600 University Street, Montreal QC, H3A 2T8, Canada}

\author{M.~P.~Rupen}
\affiliation{National Research Council of Canada, Herzberg Astronomy and Astrophysics, Dominion Radio Astrophysical Observatory, P.O. Box 248, Penticton, BC V2A 6J9, Canada}

\begin{abstract}
We present results of a search for late-time radio emission and Fast Radio Bursts (FRBs) from a sample of type-I superluminous supernovae (SLSNe-I). We used the Karl G.~Jansky Very Large Array to observe ten SLSN-I more than 5 years old at a frequency of 3~GHz. We searched fast-sampled visibilities for FRBs and used the same data to perform a deep imaging search for late-time radio emission expected in models of magnetar-powered supernovae. No FRBs were found. One SLSN-I, PTF10hgi, is detected in deep imaging, corresponding to a luminosity of $1.2\times10^{28}$\ erg s$^{-1}$. This luminosity, considered with the recent 6~GHz detection of PTF10hgi in \citet{2019arXiv190110479E}, supports the interpretation that it is powered by a young, fast-spinning ($\sim$ ms spin period) magnetar with $\sim$ 15 $M_{\sun}$ of partially ionized ejecta. Broadly, our observations are most consistent with SLSNe-I being powered by neutron stars with fast spin periods, although most require more free-free absorption than is inferred for PTF10hgi. We predict that radio observations at higher frequencies or in the near future will detect these systems and begin constraining properties of the young pulsars and their birth environments.
\end{abstract}


\section{Introduction}
\label{sec:intro}
The advent of wide-field surveys focused on the time domain has led to the discovery and characterization of new, rare classes of transient astrophysical phenomena. Optical surveys have identified extremely luminous classes of transients called superluminous supernovae \citep[SLSNe;][]{2012Sci...337..927G}. The hydrogen-poor subset of SLSNe (``type-I'') are unlikely to be powered by interaction with their circumburst medium. This suggests that something powers them internally, such as an accreting black hole \citep{2006ARA&A..44..507W} or rapidly spinning young neutron star \citep{2016MNRAS.461.1498M, 2017ApJ...841...14M}.

At cm-wavelengths, radio surveys have identified the Fast Radio Burst \citep[FRB;][]{2019A&ARv..27....4P,2019arXiv190605878C}, a coherent, millisecond transient. The recent association of an FRB with a galaxy at $z=0.1927$\ confirmed that they are extremely bright and luminous \citep{2017Natur.541...58C,OPT}, motivating new models for FRB origin \citep{2017ApJ...839L...3K}. A new suite of FRB origin models has already been published \citep{2019MNRAS.485.4091M}. However, only one model has successfully \emph{predicted} the properties of FRB 121102: young magnetars \citep{2016MNRAS.461.1498M}. 

Newborn magnetars have emerged as a strong candidate for producing a variety of luminous transients \citep[e.g.][]{2007ApJ...666.1069M,2015MNRAS.454.3311M,2015MNRAS.452.3869N,2016ApJ...818...94K,2018MNRAS.tmp.2301M}. Classes of object such as SLSNe-I, FRBs, and even ultralong gamma-ray bursts (GRBs) have severe energetic requirements that can be met by tapping into the spin-down power of a magnetar with a millisecond rotation period. The magnetar birth scenario presents a testable hypothesis: SLSNe-I should be associated with luminous pulsar wind nebulae (PWNe) at late times  \citep{2016MNRAS.461.1498M,2018MNRAS.474..573O,2017ApJ...841...14M}. It is also possible that SLSNe-I leave compact remnants that emit coherent radio emission detectable as FRBs. Coherent radio emission from pulsars is observationally well characterized and the fraction of sources detectable by this emission is roughly 10\% \citep{1998MNRAS.298..625T}.

\citet{2019arXiv190110479E} found the first observational support for the magnetar-powered supernova model with the detection of late-time radio emission coincident with the SLSN-I known as PTF10hgi. The radio source is located in a dwarf galaxy, similar to that seen for most SLSNe-I \citep{2014ApJ...787..138L, 2014ApJ...797...24V}, but it could also potentially be associated with an AGN. It is also possible that the emission is associated with the afterglow of an off-axis jet of a GRB. New observations to constrain the temporal and spectral evolution of the source will help distinguish between these classes of object. Late radio observations have been a powerful tool for studying long gamma-ray bursts \citep{2004ApJ...607L..13S}, short gamma-ray bursts \citep{2014MNRAS.437.1821M}, and tidal disruption events \citep{2013ApJ...763...84B}

Here we present a multifaceted search for signatures of magnetar birth in SLSNe-I. We use the Karl G.~Jansky Very Large Array (VLA) to search for late-time radio emission at 3~GHz that is coincident with known SLSNe-I and use the results for detailed modeling of magnetar birth models. We detect one of the ten sources, PTF10hgi, confirming work presented in \citet{2019arXiv190110479E}. We used the real-time transient search system known as \emph{realfast} to commensally search for FRBs in the same data through millisecond imaging \citep{2018ApJS..236....8L}.

\section{Data and Analysis}

\subsection{Observations}
\label{sec:obs}

We used the VLA to observe a sample of ten SLSN-I at 3 GHz. We selected the oldest ten SLSN-I from the first large sample with well-characterized host galaxies \citep{2014ApJ...787..138L}. This sample has rest-frame ages greater than 5 years, but excludes SCP 06F6, as it is predicted to be too faint to detect in a reasonable amount of time. Table \ref{tab:slsn} lists the SLSN-I in order of their rest-frame age at time of observation in late 2017. The VLA observations were designed with two goals: search for late-time radio emission and search for FRBs. The late-time radio emission from magnetar-powered supernovae is expected to fade as $t^{-2}$, but is also subject to free-free absorption by the supernova ejecta at early times \citep[e.g.][]{2017ApJ...839L...3K}. The balance of these two effects favors observations at frequencies from 2--10 GHz on timescales of 5--20 years. For the FRB search, we favored observing frequencies $\lesssim3$~GHz, where most FRBs have been observed. At lower frequencies, the VLA has a larger field of view, which also improves the odds of detecting an FRB that is unassociated with the SLSN-I.

\begin{table}[tb]
\caption{SLSN-I Sample}
\centering
\begin{tabular}{lcccccc}
\hline
Name & Redshift & R.A. & Decl. & Age &  \\
 & & (J2000) & (J2000) & (yr)  \\ \hline
SN 2005ap\tablenotemark{a}  & 0.283 & 13:01:14:83 & +27:43:32:3 & 9.9 \\
SN 2007bi & 0.127 & 13:19:20:14 & +08:55:43:7 & 9.4 \\
SN 2006oz & 0.396 & 22:08:53:56 & +00:53:50:4 & 8.0 \\
PTF10hgi\tablenotemark{c} & 0.098 & 16:37:47:04 & +06:12:32:3 & 6.8 \\
PTF09cnd & 0.258 & 16:12:08:94 & +51:29:16:1 & 6.6 \\
SN 2010kd & 0.101 & 12:08:00:89 & +49:13:32:9 & 6.4 \\
SN 2010gx\tablenotemark{b} & 0.23 & 11:25:46:71 & -08:49:41:4 & 6.2 \\
PTF09cwl & 0.349 & 14:49:10:08 & +29:25:11:4 &  6.1 \\
SN 2011ke & 0.143 & 13:50:57:77 & +26:16:42:8 & 5.7 \\
PTF09atu & 0.501 & 16:30:24:55 & +23:38:25:0 & 5.5 \\
\end{tabular}
\tablenotetext{a}{Late-time radio limit at 1.4~GHz by \citet{2018MNRAS.473.1258S}.}
\tablenotetext{b}{Late-time radio limit at 3~GHz by \citet{2018ApJ...857...72H}.}
\tablenotetext{c}{Late-time radio detection at 6~GHz by \citet{2019arXiv190110479E}.}
\label{tab:slsn}
\end{table} 

We observed with the 3~GHz band as a compromise between the expected late-time emission and FRB detection goals. We used 8 spectral windows covering the full frequency range from 2.5--3.5~GHz using 32 channels per window with a width of 4~MHz per channel. The visibility data were recorded with 5~ms cadence \citep[comparable to FRB pulse width;][]{frbcat} to allow a real-time search for FRBs with \emph{realfast}\footnote{See also: \url{http://realfast.io}}. The antennas were in the ``B'' configuration, which has baseline lengths up to 10 km and a synthesized beam size of roughly 3\arcsec\ at 3~GHz. These data are thus sensitive to FRBs anywhere within the primary beam, which has a full-width-at-half-power of 14\arcmin.

Table \ref{tab:obs} describes the observations of each target. The ten targets were scheduled in four groups, each of which was observed in two epochs from late 2017 to early 2018. The observing duration for each epoch was set to detect a source with a power roughly 10 times lower than the persistent radio source associated with FRB 121102 ($3\sigma$ power sensitivity of $L_{3\rm{GHz}}=3\times10^{28}$\ erg s$^{-1}$ Hz$^{-1}$).

For three of the observing epochs (MJD 58128, 58130, and 58131), the correlator was not able to write data fast enough, so some data were lost. Roughly $\sim20$\% of data were affected by correlator issues, interference, or bad calibration solutions; in some later observations, up to 50\% of data were lost.

\subsection{Fast transient search}
\label{sec:fastsearch}

After each observation, we searched the 5~ms data for FRBs with the \textit{rfpipe} search pipeline \citep{2017ascl.soft10002L}. The search was run offline using CPUs in spare nodes of the VLA correlator cluster. This search applies calibration solutions calculated in realtime by the VLA observing system (a.k.a. \textit{telcal}). Bad channels and integrations are flagged using a sigma clipping algorithm, while the variance of visibilities over baselines is used to flag near-field interference for specific channel-integration-polarization bins.

We searched for FRBs with dispersion measures up to 3000 pc cm$^{-3}$ and pulse widths up to 40~ms. The maximum distance for this SLSN-I sample is $z\sim0.5$\, which implies DM contribution from the intergalactic medium of roughly 400 pc cm$^{-3}$ \citep{2019MNRAS.485..648P}. The DM contribution from the Milky Way is smaller than the extragalactic contribution in all cases \citep{2002astro.ph..7156C}. The DM contribution from the FRB environment and host galaxy is generally expected to be less than 3000 pc cm$^{-3}$ \citep{2014ApJ...797...70K}; in the case of FRB 121102, this component contributes less than 225 pc cm$^{-3}$ to the total DM measurement \citep{OPT}.

All candidates brighter than $8\sigma$ were inspected by looking at dedispersed burst spectra and 5~ms image associated with the event. No bursts were found brighter than $8\sigma$. A typical observation had 26 antennas and 1.5~GHz of clean bandwidth, which corresponds to a sensitivity of roughly 4~mJy per 5~ms snapshot image.

The nominal sensitivity is idealized and needs to be corrected for the effects of dedispersion and primary beam attenuation. The \textit{rfpipe} search uses a brute-force dedispersion algorithm that can lose sensitivity to pulses with DM between the DM search grid \citep{2015MNRAS.447.2852K}. The DM search grid was set to lose at most 5\% of the nominal sensitivity due to intra-DM sensitivity losses, so the $8\sigma$ limit is thus 34 mJy in 5 ms at the center of the primary beam. The image search was also sensitive to FRBs throughout the primary beam, which has a FWHM of 14\arcmin at 3~GHz. The search for FRBs throughout the primary beam was complete to a flux limit of 68 mJy in 5 ms. The sensitivity is best defined as a fluence limit averaged over the observing band from 2.5--3.5~GHz, so sensitivity to temporally or spectrally narrow emission structure is worse than stated here \citep{2017ApJ...850...76L,2018ApJ...863....2G}.

\subsection{Deep imaging}
\label{sec:deepimaging}

We averaged the 5-ms integrations to 1~s and analyzed these new data sets with the CASA calibration pipeline \citep{2007ASPC..376..127M}. Any 1~s integration with more than 30\% of its subintegrations flagged was fully flagged. 

Three of the observing blocks (including seven of the targets) used a standard flux calibrator (either 3C286 or 3C48). These fields were calibrated with the VLA CASA calibration pipeline (version 5.4.0). One observing block, including PTF09atu, PTF09cnd, and PTF10hgi, used 3C295 as a flux calibrator, which is not supported by the latest pipeline. For these observations, we instead used the VLA scripted pipeline (version 1.4.0). In all cases, calibration quality was validated by inspecting the standard pipeline output of calibrator images, gain solutions, and visibility plots.

Both epochs of all ten fields were imaged with \textit{tclean} in CASA. For each field, we first produced a sky model through a light clean of mJy-brightness sources using natural weighting. In some cases, that model was sufficient to self-calibrate the field at both epochs with a single solution per antenna and spectral window (Stokes I). We then created a final map for each field by combining both epochs and creating a deeply cleaned image. For images with image artifacts from nearby sources, we use robust weighting of 0.5. The best images from either natural or robust weighting were used to estimate noise and search for radio emission from the SLSN.

\begin{table}[tb]
\caption{Observations of SLSN-I}
\centering
\begin{tabular}{lccc}
\hline
Name & Epochs & Obs.\ Time & Sensitivity \\
 & (MJD) & (min; total) & ($\mu$Jy beam$^{-1}$; 1$\sigma$) \\ \hline
SN 2005ap & 58060, 58131 & 57 & 10 \\
SN 2007bi & 58074, 58128 & 34 & 22 \\
SN 2006oz & 58036, 58124 & 60 & 8 \\
PTF10hgi\tablenotemark{a} & 58045, 58130 & 26 & 14 \\
PTF09cnd & 58045, 58130 & 46 & 11 \\
SN 2010kd & 58074, 58128 & 27 & 14 \\
SN 2010gx & 58074, 58128 & 41 & 11 \\
PTF09cwl & 58060, 58131 & 73 & 9 \\
SN 2011ke & 58060, 58131 & 35 & 12 \\
PTF09atu & 58045, 58130 & 109 & 8 \\
\end{tabular}
\tablenotetext{a}{Detection with peak flux density of 47 $\mu$Jy.}
\label{tab:obs}
\end{table} 

Table \ref{tab:obs} lists the measured sensitivity of a deep image made for each SLSN. Only one of the targets, PTF10hgi is detected with greater than $3\sigma$\ significance. For three of the fields (PTF09cnd, SN2007bi, SN2006oz), we detected a radio source within 1\arcmin\ of the SLSN. However, all of these radio sources are offset by more than 10\arcsec (far larger than any astrometric uncertainty), which makes them highly unlikely to be associated with the SLSN or their host galaxies \cite[host galaxy images at ][]{2014ApJ...787..138L}. Stacking all ten images by the inverse noise squared gives an image with no significant source at the location of the SLSN-I and a $3\sigma$\ limit of $8\times10^{-7}$~Jy.

\begin{figure}[tb]
    \centering
    \includegraphics[width=\columnwidth]{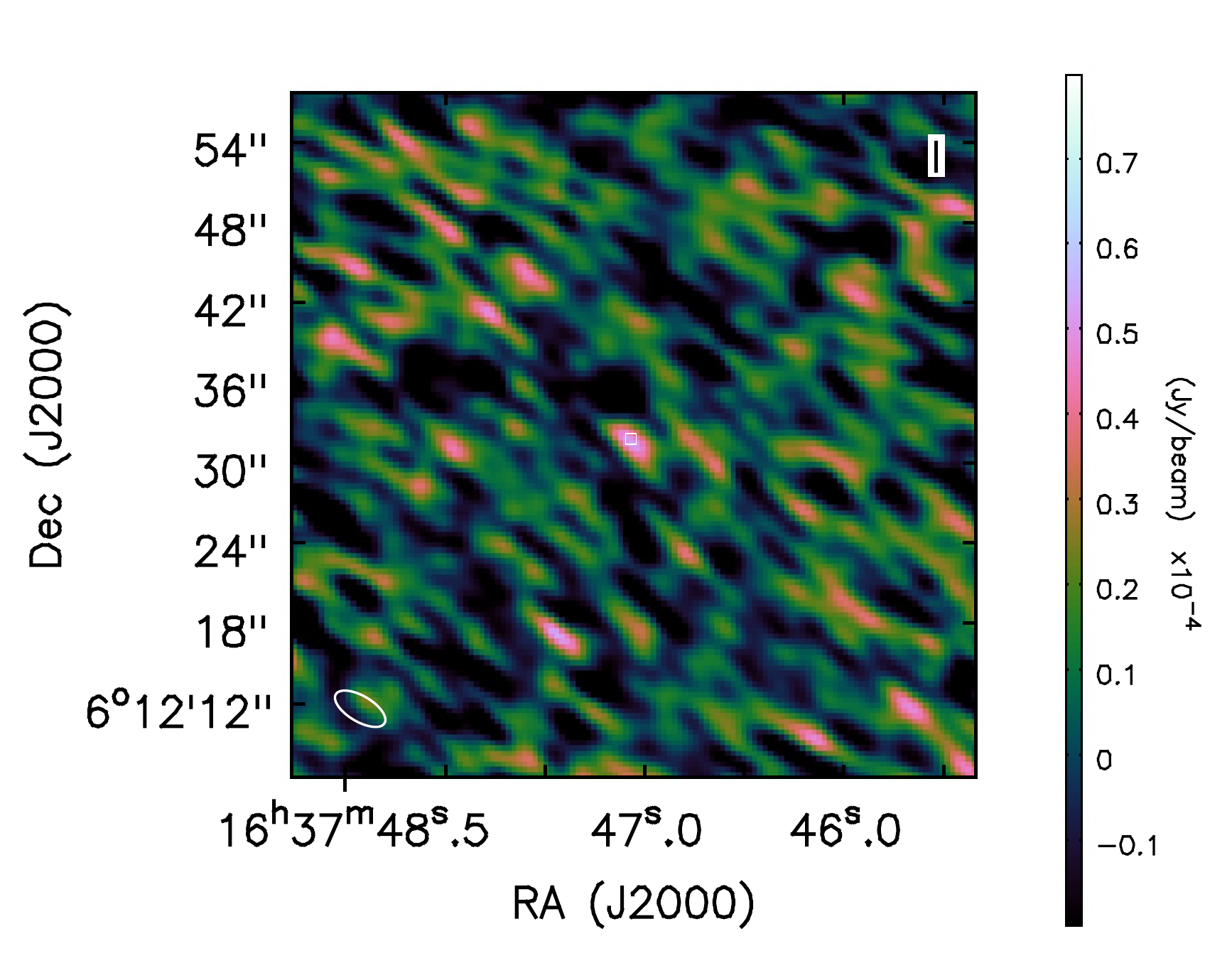}
    \caption{VLA 3~GHz radio image of PTF10hgi. The location of the 6 GHz source associated with PTF10hgi is shown with a white square \citep{2019arXiv190110479E}.}
    \label{fig:ptf10hgi}
\end{figure}

Figure \ref{fig:ptf10hgi} shows a compact source at the location of the 6~GHz counterpart to PTF10hgi \citep{2019arXiv190110479E}. This source is apparent in both observing epochs and is robust to a range of assumptions for imaging and self-calibration parameters. We modeled this region with the CASA \textit{imfit} tool to define a 2d Gaussian with width fixed to the synthesized beam shape. The best fit source has a peak flux density of $47\pm14~\mu$Jy located at (RA, Dec) (J2000) $=$\ (16:37:47.04, 6:12:31.4) with centroid uncertainty of $0.7$\arcsec$\times0.4$\arcsec. The 3~GHz radio source is coincident with the 6~GHz source located at (16:37:47.071, 6:12:31.88). A $3\sigma$\ detection significance corresponds to a false alarm rate of $10^{-3}$ or a 1\% probability of detection in a sample of ten sources. Using the observed noise properties of the PTF10hgi 3~GHz image, we estimate a smaller than 1 in 100 chance of false association with this source.

The 6~GHz source is consistent with the optical position of PTF10hgi \citep{2014ApJ...787..138L}, but the 3~GHz location is offset roughly 1\arcsec\ from optical position. We attribute this to small phase calibration errors, which can affect localizations smaller than the 3\arcsec\ synthesized beam size. Uncertainties in referencing the optical and radio frames may also contribute. Hereafter, we assume that the 3~GHz and 6~GHz sources are coincident with each other and the SLSN-I.

The 3~GHz flux density of PTF10hgi corresponds to a luminosity L$_\nu=1.2\pm0.4\times10^{28}$\ erg s$^{-1}$\ Hz$^{-1}$. Observations of PTF10hgi at 3 and 6~GHz were made within three months of each other, so they are effectively simultaneous in the context of synchrotron emission models (see \S \ref{sec:disc}). Comparing this 3~GHz flux density to the 6~GHz measurement implies a spectral index $\alpha=0.0\pm{0.6}$ (F$_\nu \propto \nu^\alpha$). This spectral index measurement is consistent with, and slightly more precise than, that of \citet{2019arXiv190110479E}.

\section{Discussion}
\label{sec:disc}

This study is the first search for late-time radio emission and FRBs from a sample of SLSN-I. There are only three SLSN-I with prior observational constraints on late-time ($>5$~yr) radio emission. \citet{2018MNRAS.473.1258S} present an upper limit of $F_{1.4~\rm{GHz}}<75~\mu$Jy\ (3$\sigma$) for SN2005ap roughly 10 years after explosion, \citet{2018ApJ...857...72H} present 3 GHz upper limits on 8 SLSNe (5 of which were SLSNe-I, one of which is in our sample), and \citet{2019arXiv190110479E} detect PTF10hgi with a flux of $F_{6~\rm{GHz}}=47.3\pm7.1~\mu$Jy roughly 7 years after explosion.



\subsection{Modelling}


In a magnetar-powered supernova, the persistent radio source luminosity is defined by the magnetar-driven wind interacting with its surrounding supernova remnant \citep{2016MNRAS.461.1498M,2017ApJ...841...14M,2018MNRAS.474..573O}. The magnetar birth properties (especially initial spin period and magnetic field strength) are inferred from the early optical light curve. We use optical data from the Open Supernova catalog \footnote{https://sne.space/} \citep{2017ApJ...835...64G} and fit by eye with a 3-parameter model \citep{2016ApJ...818...94K}: the initial spin period $P$ and the magnetic field $B_{13} = B$/($10^{13}$ G) of the neutron star, and the mass $M_{\rm ej}$ of the supernova ejecta. Using this method, we can determine the parameters to within 5-10\%.  Since the magnetar model has degeneracies, we define one parameter set with $P$ = 1 ms ($P_{\text{min}}$), which is close to the mass-shedding limit for neutron stars \citep{2016RvMP...88b1001W} and another with a larger period ($P_{\text{max}}$), which is the largest spin period consistent with the optical light curve.  \cite{2018MNRAS.474..573O} finds that $P_{\text{max}}$ varies between supernovae, but is typically less than 5~ms. 

There are multiple approaches to modeling optical light curves of SLSNe-I \citep[e.g.,][]{2013ApJ...770..128I, 2017MNRAS.464.3568P, 2017ApJ...842...26L, 2017ApJ...850...55N} and they sometimes derive different magnetar parameters for the same sources. Earlier studies tended to assume simple dipole spin down model, while we use a model based on numerical simulation \citep{2005PhRvL..94b1101G, 2006ApJ...648L..51S, 2013MNRAS.435L...1T}. The numerical simulations require much smaller $B_{13}$\ and $P$\ for a given spin-down luminosity \citep{2016ApJ...818...94K}. Our treatment also allows for acceleration of the ejecta due to interaction with the PWN, which couples the dynamics of the ejecta to the spin-down luminosity; a realistic ejecta profile \citep[a homologous core;][]{2010ApJ...717..245K}; and self-consistently treats the radio and optical signatures to break degeneracies inherent to the optical data alone.


Once the magnetar parameters have been found, we calculate the time evolution of the radio emission from the PWN based on these optically-derived parameters. This emission from the PWN is calculated as in previous papers~\citep[see][and references therein]{2006ARA&A..44...17G,2010ApJ...715.1248T}. We model not only the dynamics of PWNe and SNe as in our three-parameter optical model, but also self-consistently calculate pair cascades, Compton and inverse Compton scattering, adiabatic cooling and both internal and external attenuation by solving the Boltzmann equation for electron/positrons and photons in the PWN over all electron energies and photon frequencies \citep{2015ApJ...805...82M,2016MNRAS.461.1498M}. We assume an electron-positron injection spectrum motivated by Galactic PWNe such as the Crab PWNe~\citep[e.g.,][]{2010ApJ...715.1248T, 2013MNRAS.429.2945T}, a broken power law with a peak Lorentz factor of $\gamma_{b} = 10^{5}$ and injection spectral indices of $q_1=1.5$ and $q_2=2.5$. Free-free absorption in the ejecta is calculated assuming a singly ionized oxygen ejecta, and we do not consider absorption outside the ejecta. 

\subsection{PTF10hgi}
\label{sec:hgi}

Given the new, more precise, spectral index measurement, we discuss the viability of different astrophysical models for PTF10hgi. As discussed in  \citet{2019arXiv190110479E}, there are three viable models for this radio emission: an AGN, an off-axis GRB jet, and a nebula produced by a remnant magnetar. The spectral index is consistent with radio-loud AGN \citep{1994ApJS...95....1E, 2017Natur.541...58C}, but the AGN scenario is unlikely \citep{2019arXiv190110479E} because either the black hole would have an unexpectedly large mass for a radio-quiet AGN (5\% of the host galaxy, while dwarf galaxy black holes are generally $\lesssim$ 0.1\% of the total mass \citep{2003MNRAS.345.1057M, 2013ApJ...775..116R}) or the host galaxy would be peculiar, since the prevalence of radio-loud AGN in dwarf galaxies is $\lesssim$ 1\% \citep{2013ApJ...775..116R}. The off-axis GRB model predicts bright emission at earlier times, well above the limits placed on other SLSNe \citep{2018ApJ...856...56C, 2019arXiv190110479E}, and predicts a spectral index $\alpha \sim -1$, which is disfavoured by our observations.

\begin{figure*}[tb]
    \centering
    \includegraphics[width=0.45\textwidth]{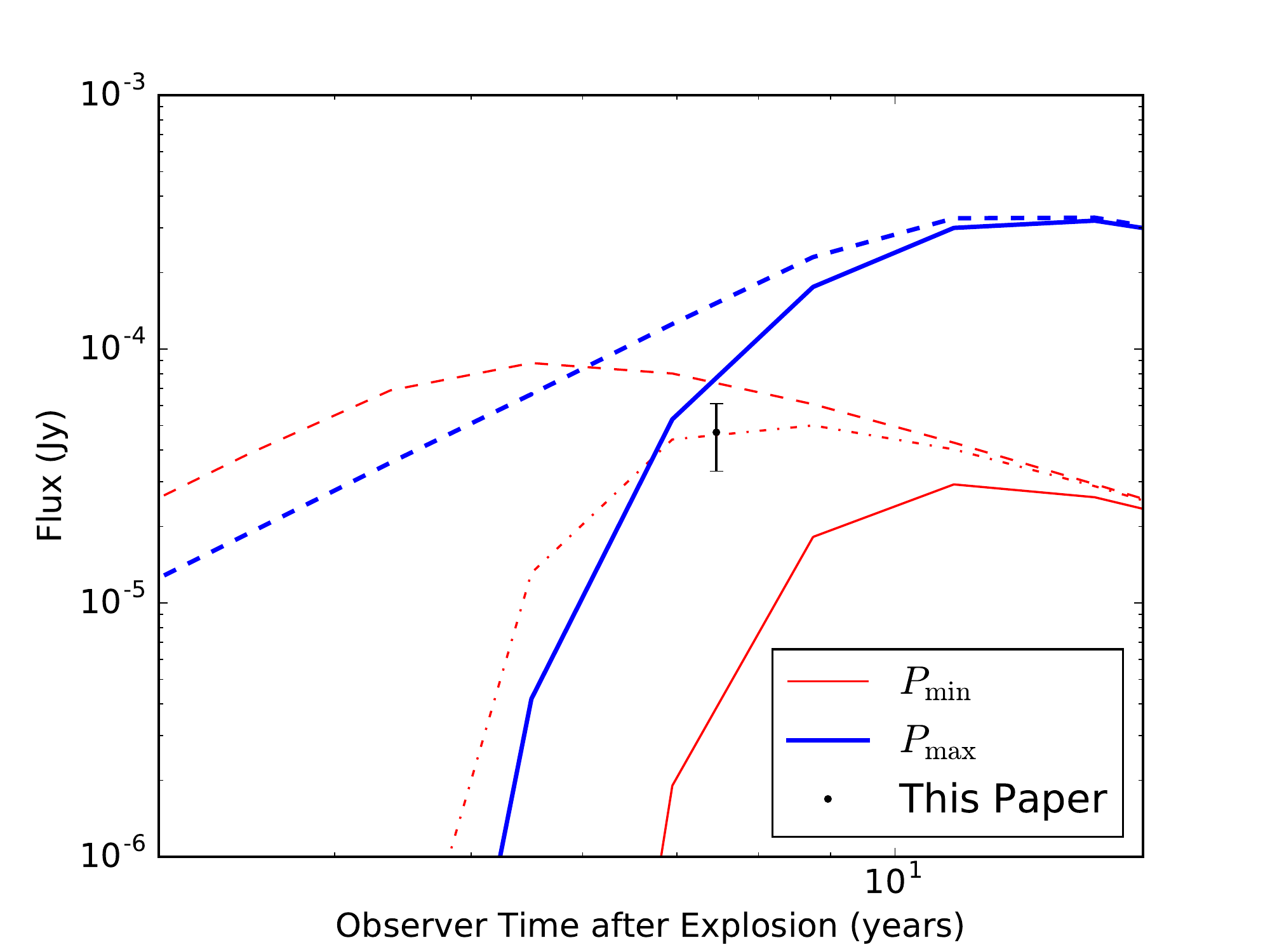}
    \includegraphics[width=0.45\textwidth]{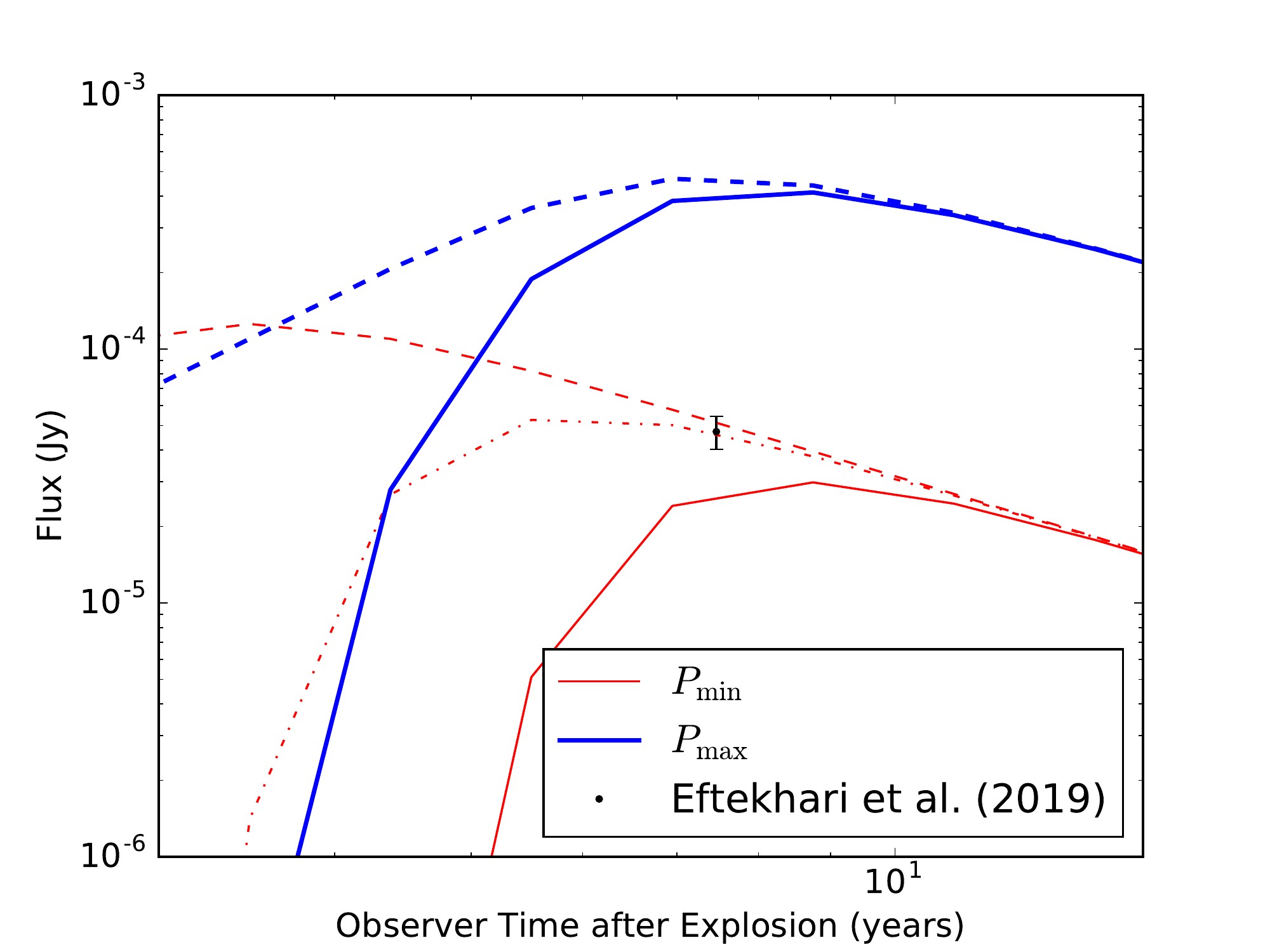}
    \caption{The detected fluxes for PTF10hgi at 3 GHz (left panel; this paper) and 6 GHz \citep[right panel; ][]{2019arXiv190110479E} with their 1$\sigma$ uncertainties shown in black.  The $P_{\text{min}}$ (1~ms) and $P_{\text{max}}$ models shown in Table \ref{tbl:snparam} are displayed in red and blue respectively, with the solid lines indicating the light curve with absorption and dashed lines indicating the curve with no absorption. We find that a $P_{\text{min}}$ model with 30-50\% of the ejecta singly ionized, with the rest neutral, can reproduce the observed data; the dash-dotted line for $P_{\text{min}}$ indicates a model with 40\% of the ejecta ionized.}
    \label{fig:ptf10}%
\end{figure*}

\begin{figure*}[tb]
    \centering
    \includegraphics[width=0.45\textwidth]{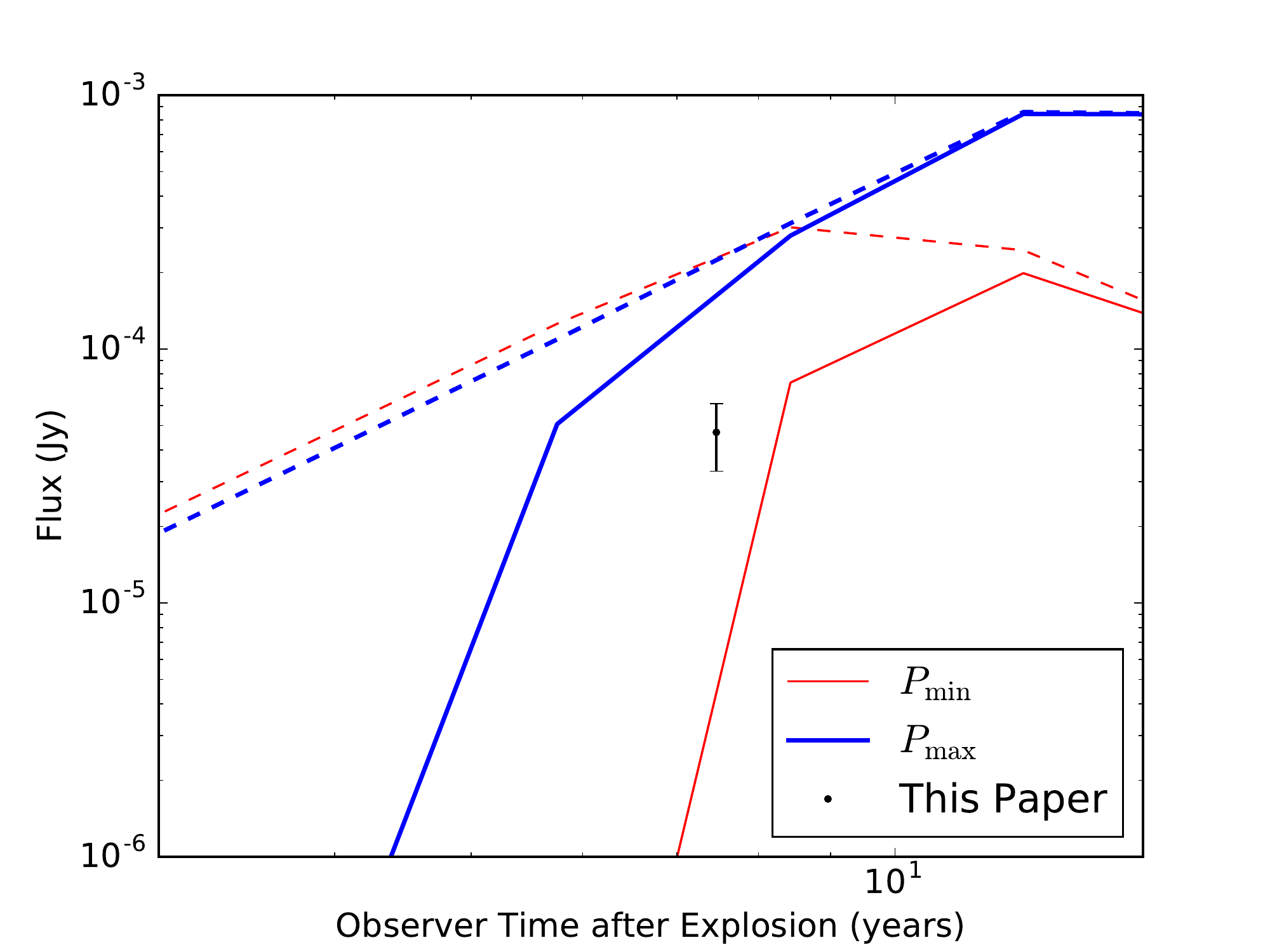}
    \includegraphics[width=0.45\textwidth]{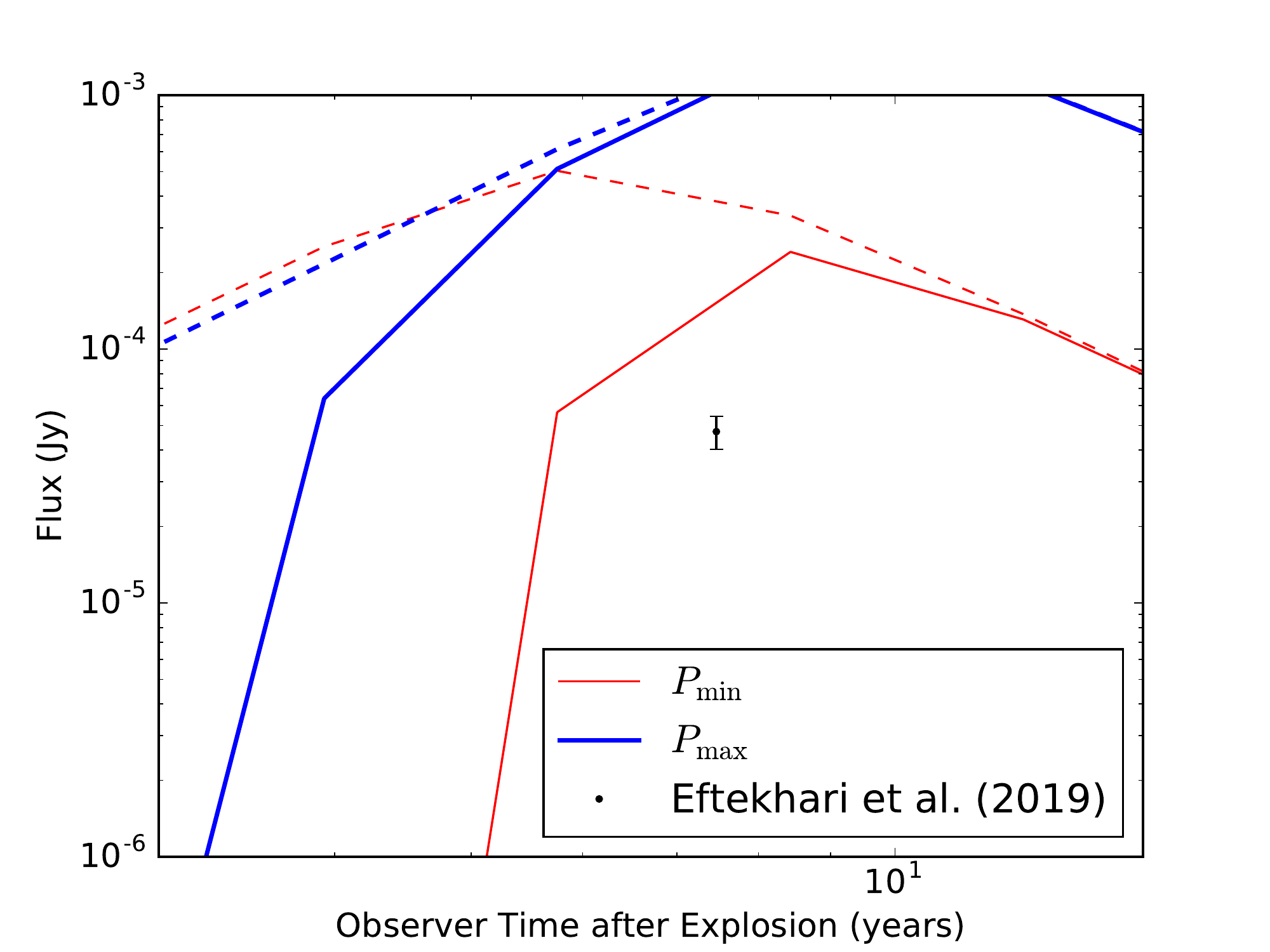}
    \caption{The same as Figure \ref{fig:ptf10}, but with $\gamma_b = 10^2$ instead of $10^5$.}
    \label{fig:ptf10gb1e2}%
\end{figure*}

The magnetar model predictions for the $P_{\text{min}}$ and $P_{\text{max}}$ cases, along with the data at 3 and 6 GHz, are shown in Figure \ref{fig:ptf10}.  The $P_{\text{max}}$ model slightly overpredicts the data at 3 GHz, but severely overpredicts at 6 GHz, while the $P_{\text{min}}$ model slightly underpredicts at both frequencies with absorption, but is close to fitting both points with little or no absorption. We find that a $P_{\text{min}}$ model with 30-50\% of the ejecta singly ionized, with the rest neutral, can reproduce the observed data; a model with 40\% ionization is shown in Figure \ref{fig:ptf10}.  This might be expected for a large ejecta mass, or the lack of absorption could mean that the ejecta are clumped in some regions away from the line of sight and relatively unobstructed along the line of sight. The $P_{\text{min}}$ model also disfavours a Wolf-Rayet progenitor, as the ejecta mass is larger than expected in that model.

We can also use this detection to constrain the electron injection spectrum.  The injection Lorentz factor $\gamma_b$, which governs the frequency of the spectral break where most of the energy is injected, can not be constrained in optical observations, as they are only sensitive to the total energy injected, and could take values from $10^2-10^6$.  Figure \ref{fig:ptf10} shows models with $\gamma_b = 10^5$, which is also assumed in \cite{2018MNRAS.474..573O}; this means the $\nu F_\nu$ synchrotron spectrum peaks at UV/x-ray energies.  Figure \ref{fig:ptf10gb1e2} shows the light curves for the same parameters, except with $\gamma_b = 10^2$, giving a spectrum that peaks at infrared/microwave energies --- models for the persistent emission from FRB 121102 usually have spectra that peak in this range \citep[][Omand et al. in prep]{2018arXiv180809969M}.  We see that the luminosity at peak in the radio bands is much higher, completely excluding all models at 6 GHz, even though the light curve has yet to reach its peak.  Based on this result, we show that these results favour higher values of $\gamma_b$, and exclude those with $\gamma_b \lesssim 10^4.$

Overall, the detections in both bands are most consistent with the magnetar model, but observations at more frequencies and epochs will be needed to determine the system properties with any certainty. Models involving fast cooling emission predict a strong evolution in the spectral index through its peak luminosity, up until there is a consistent, negative spectral index ($\alpha=-1/2$; Omand et al. in prep). Models involving relic cooling emission predict a weak evolution of the spectral index, remaining almost flat even after peak \citep[e.g.][]{2018MNRAS.474..573O, 2018arXiv180809969M}. The measured flat spectral index suggests that PTF10hgi may be near peak and disfavours a detection early in the rise of the radio emission, but only further observations can differentiate the two scenarios.

\begin{figure*}[ht]
    \centering
    \includegraphics[width=0.33\textwidth]{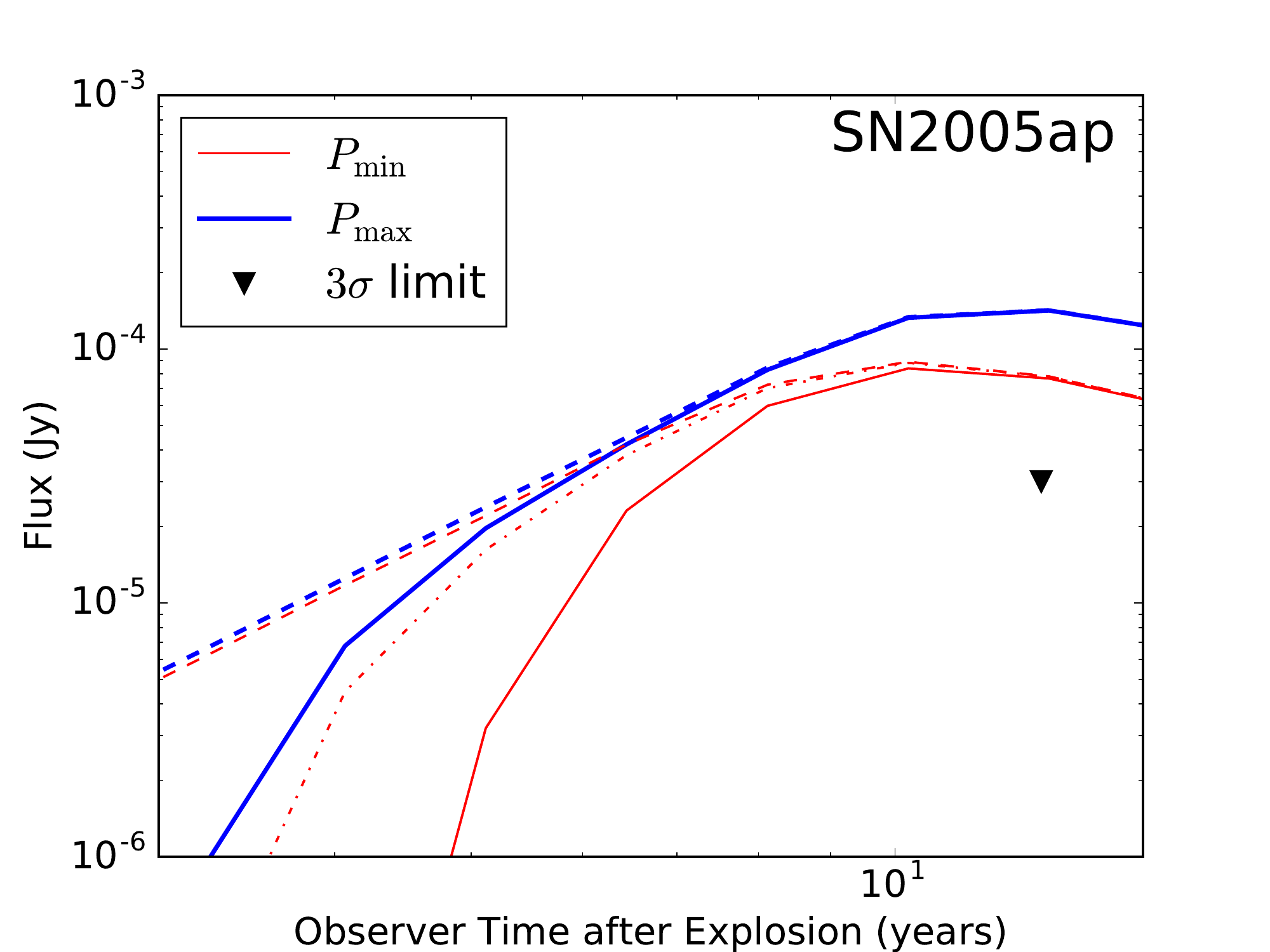}
    \includegraphics[width=0.33\textwidth]{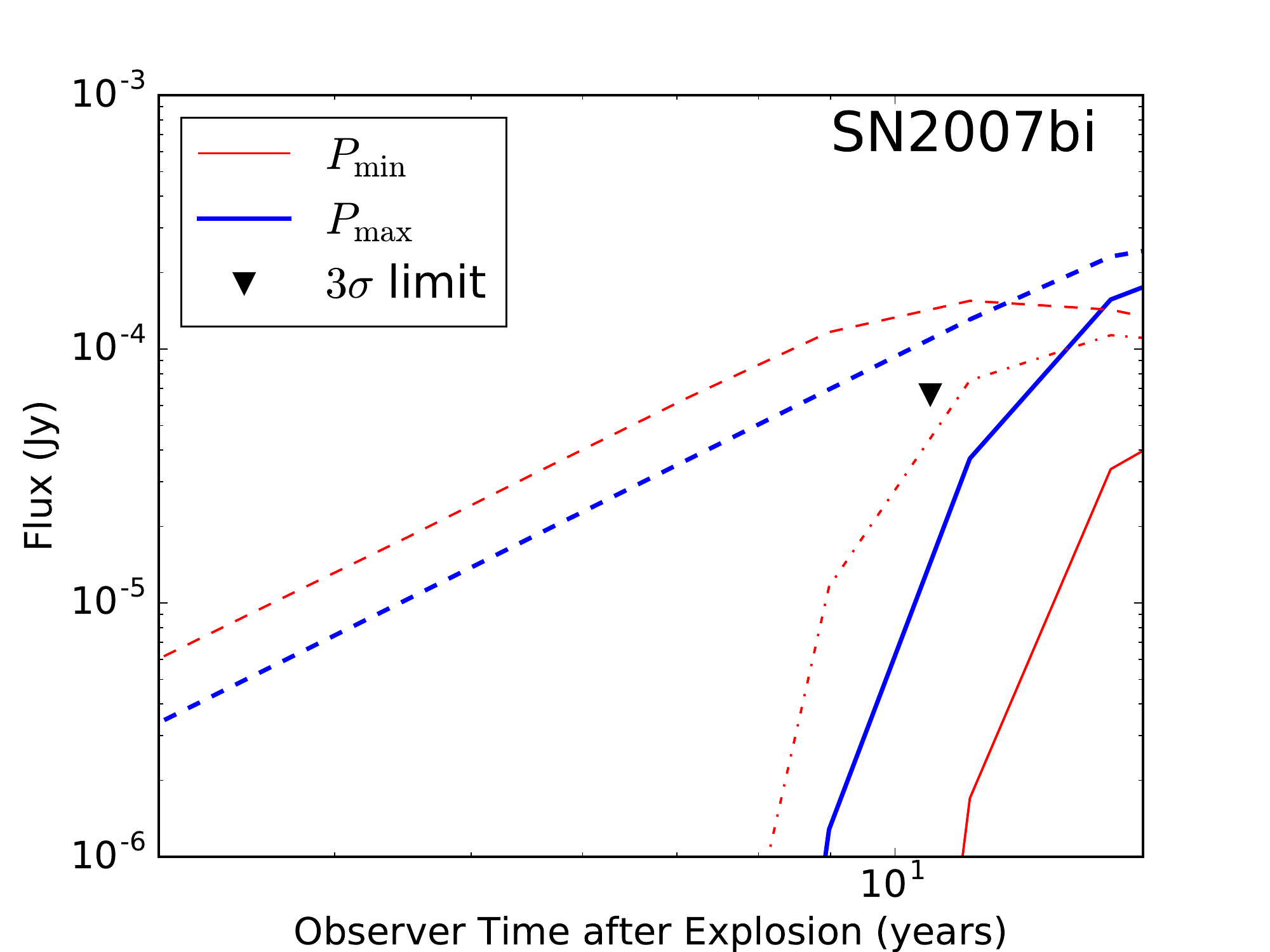}

    \includegraphics[width=0.33\textwidth]{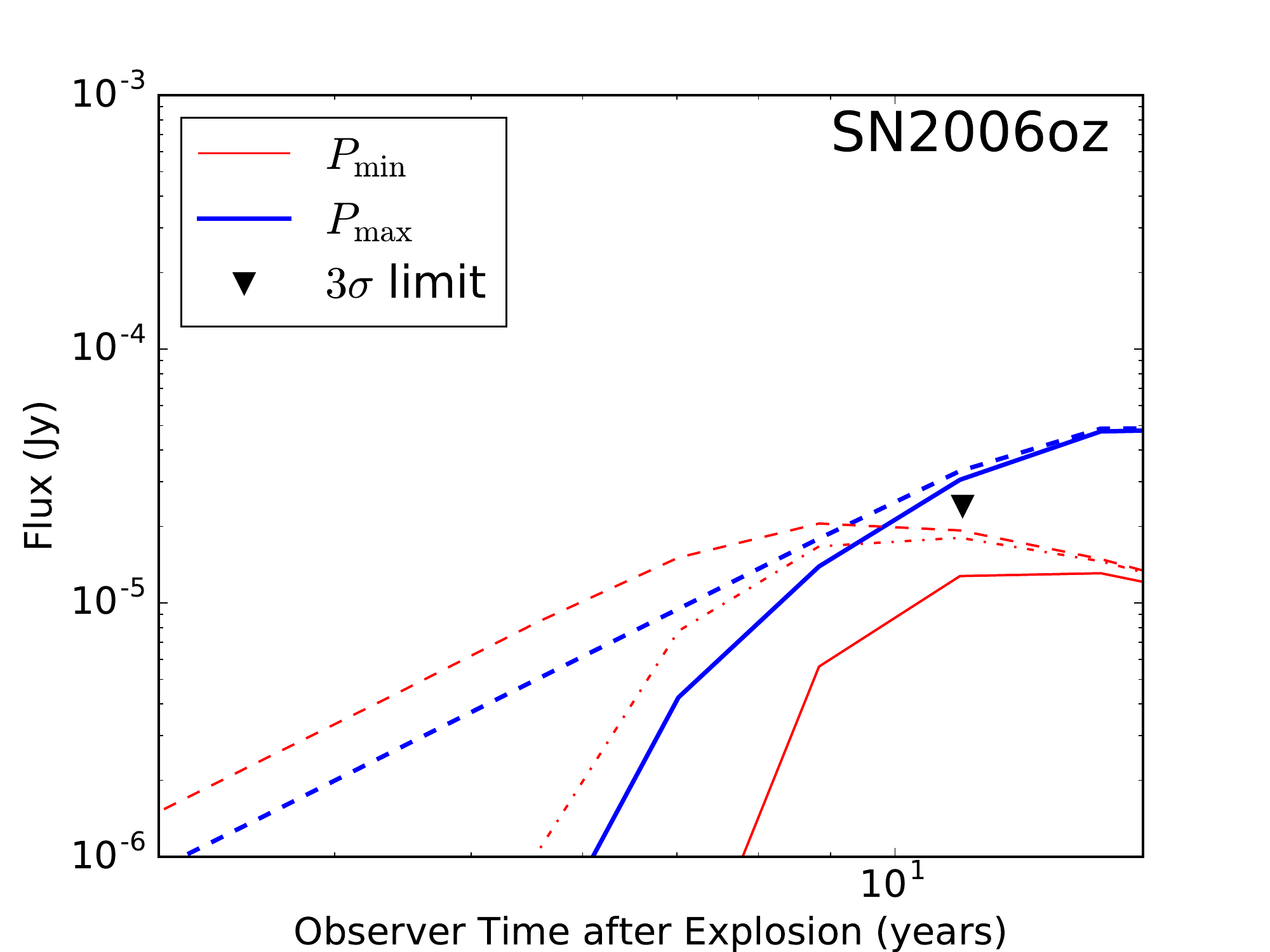}
    \includegraphics[width=0.33\textwidth]{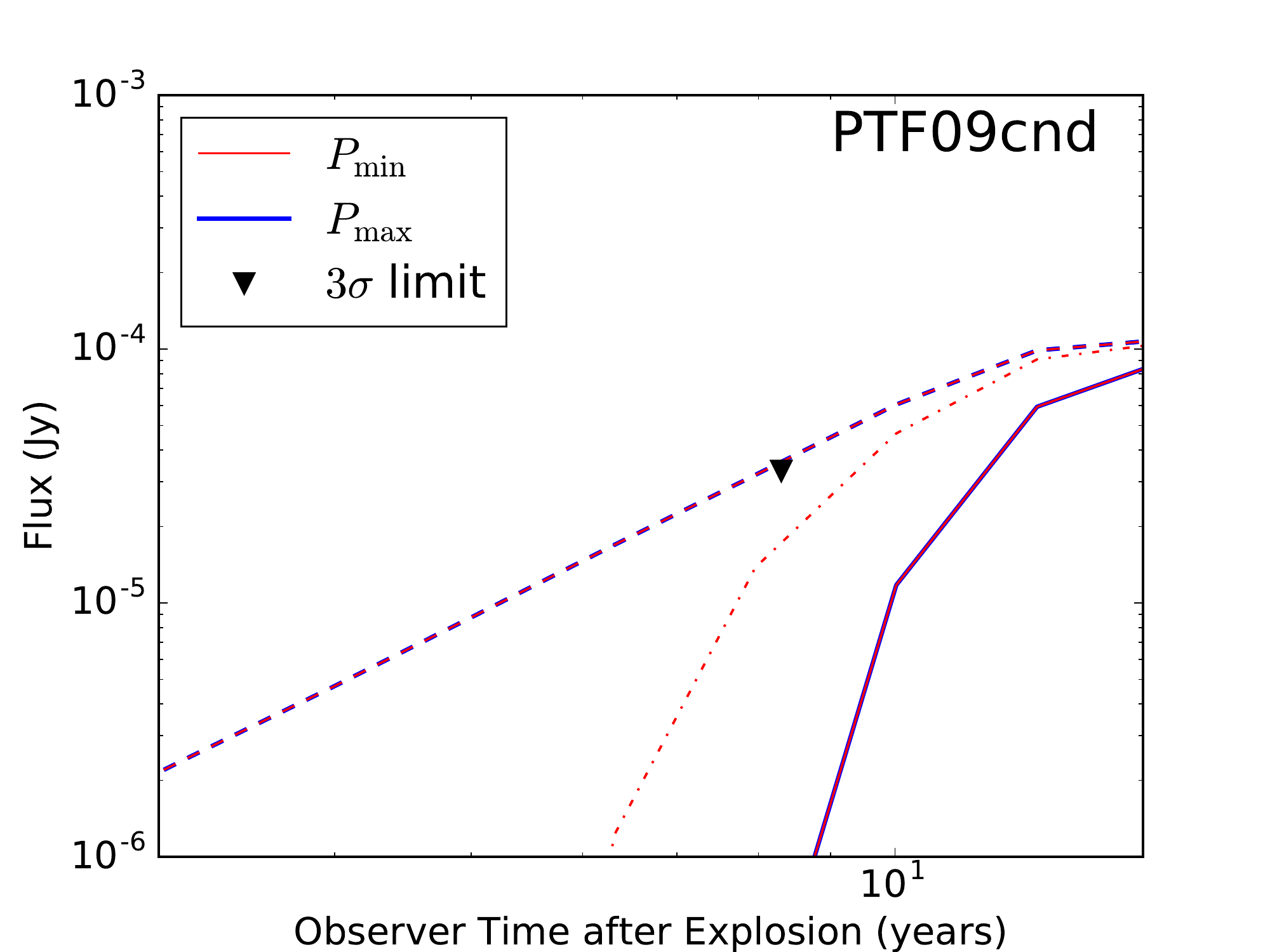}

    \includegraphics[width=0.33\textwidth]{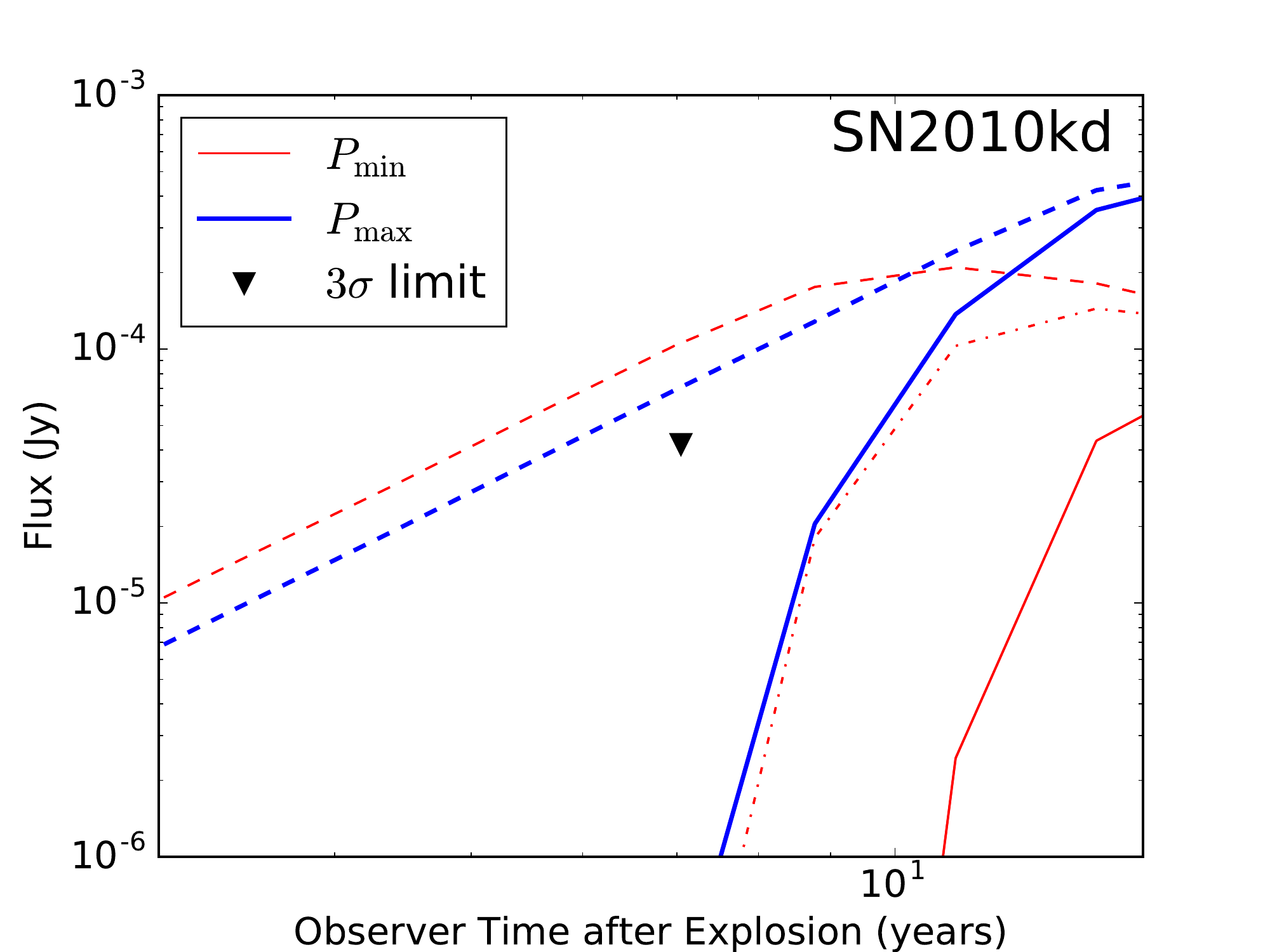}
    \includegraphics[width=0.33\textwidth]{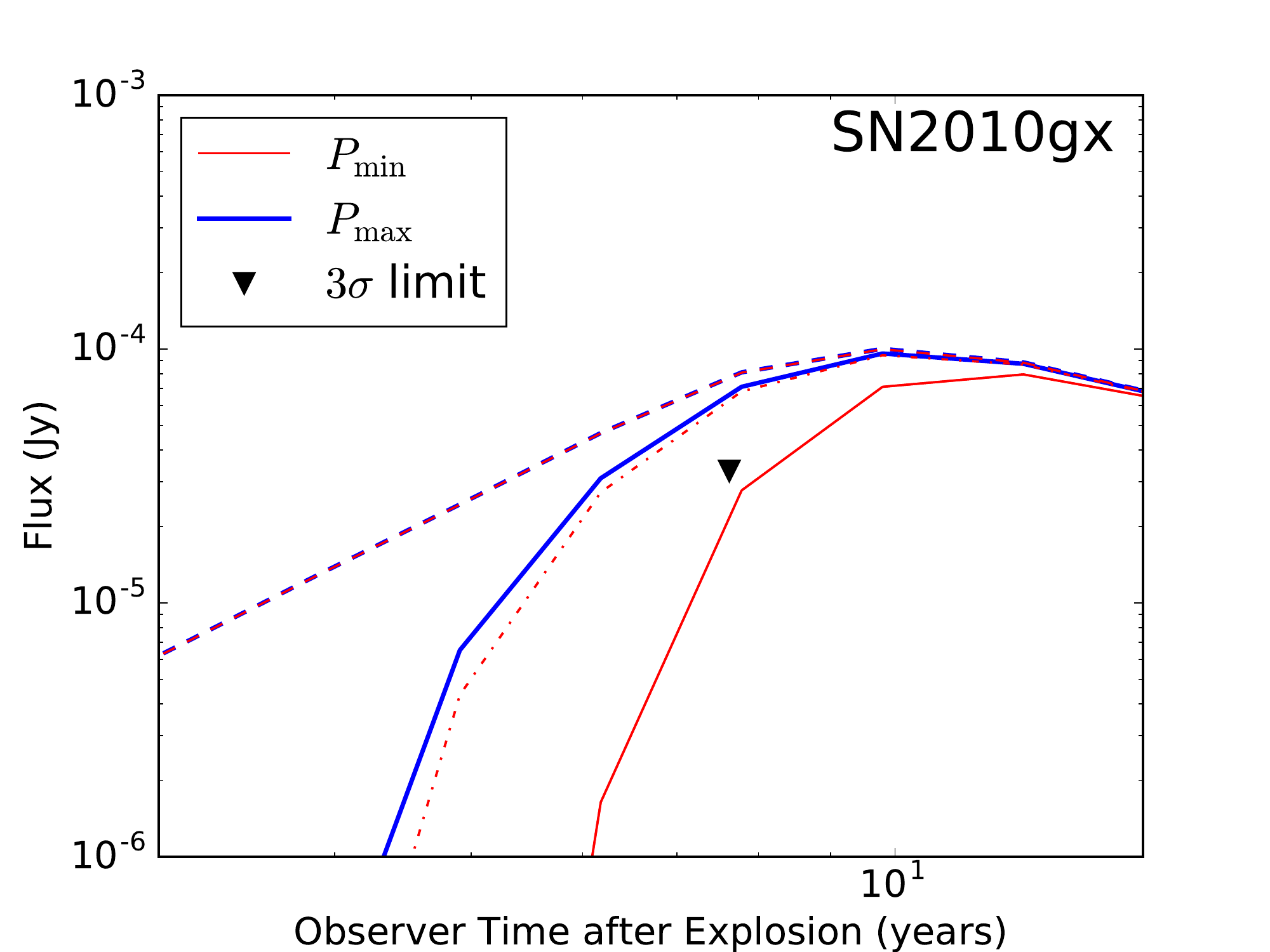}

    \includegraphics[width=0.33\textwidth]{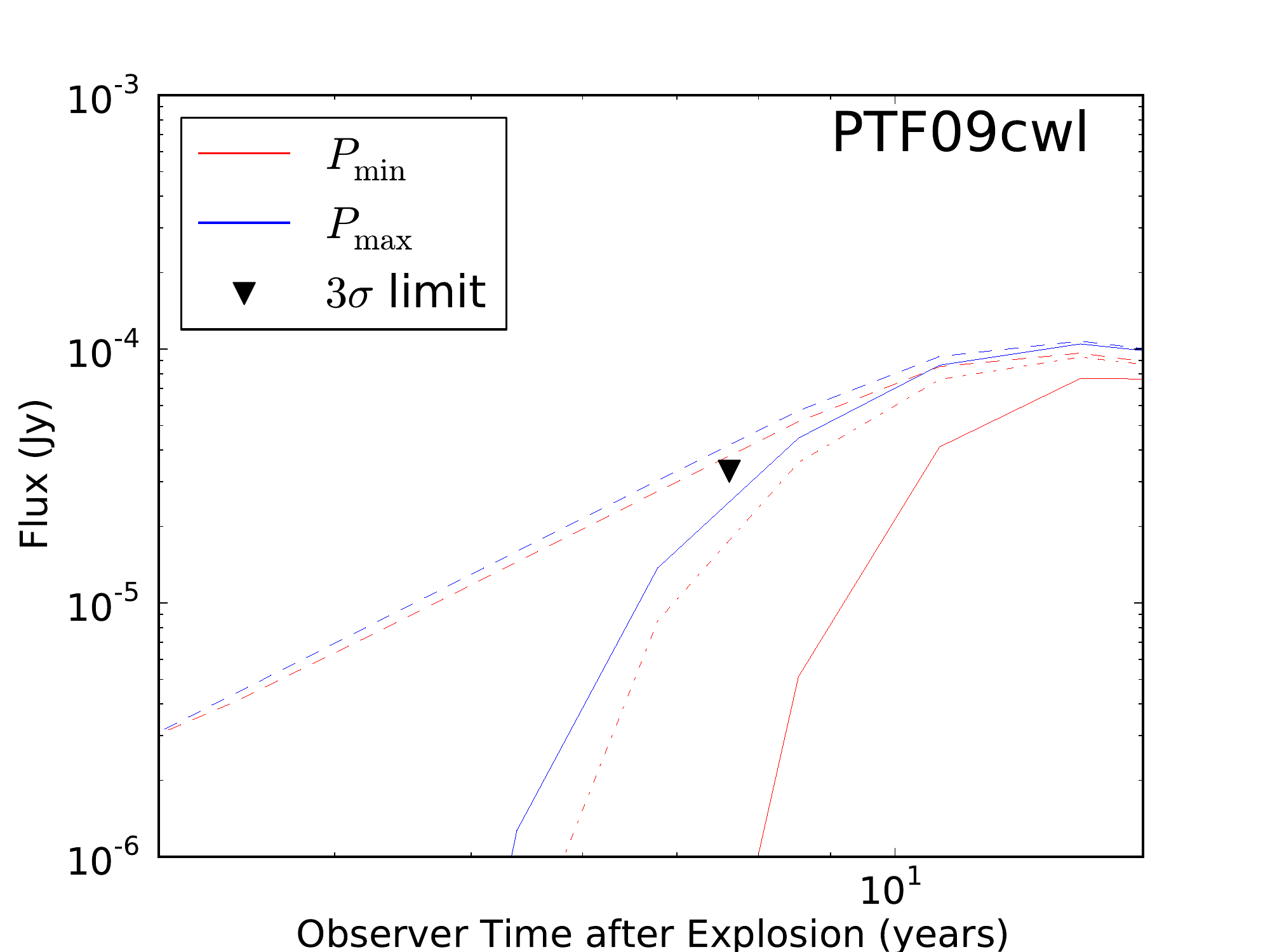}
    \includegraphics[width=0.33\textwidth]{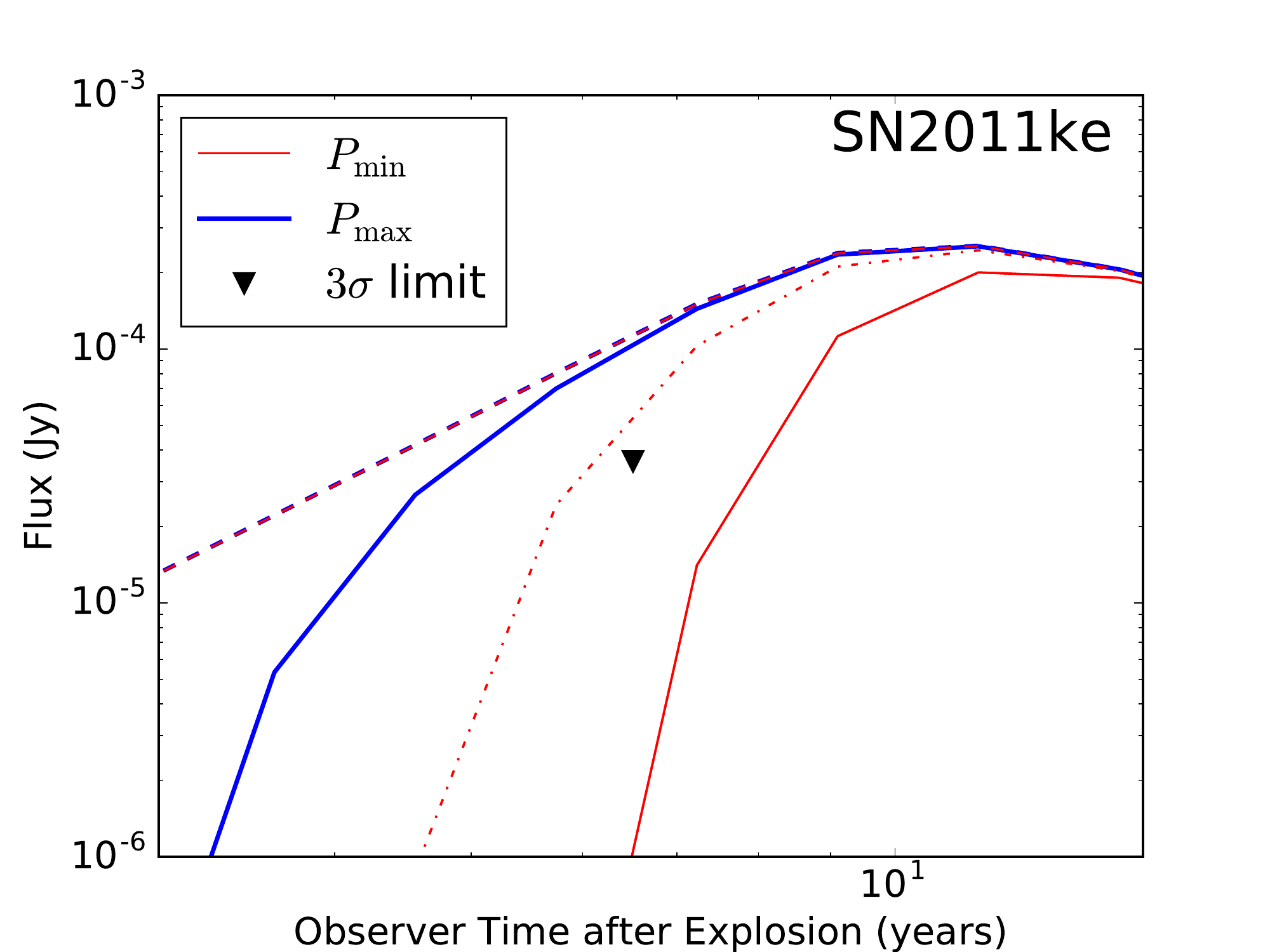}
    
    \includegraphics[width=0.33\textwidth]{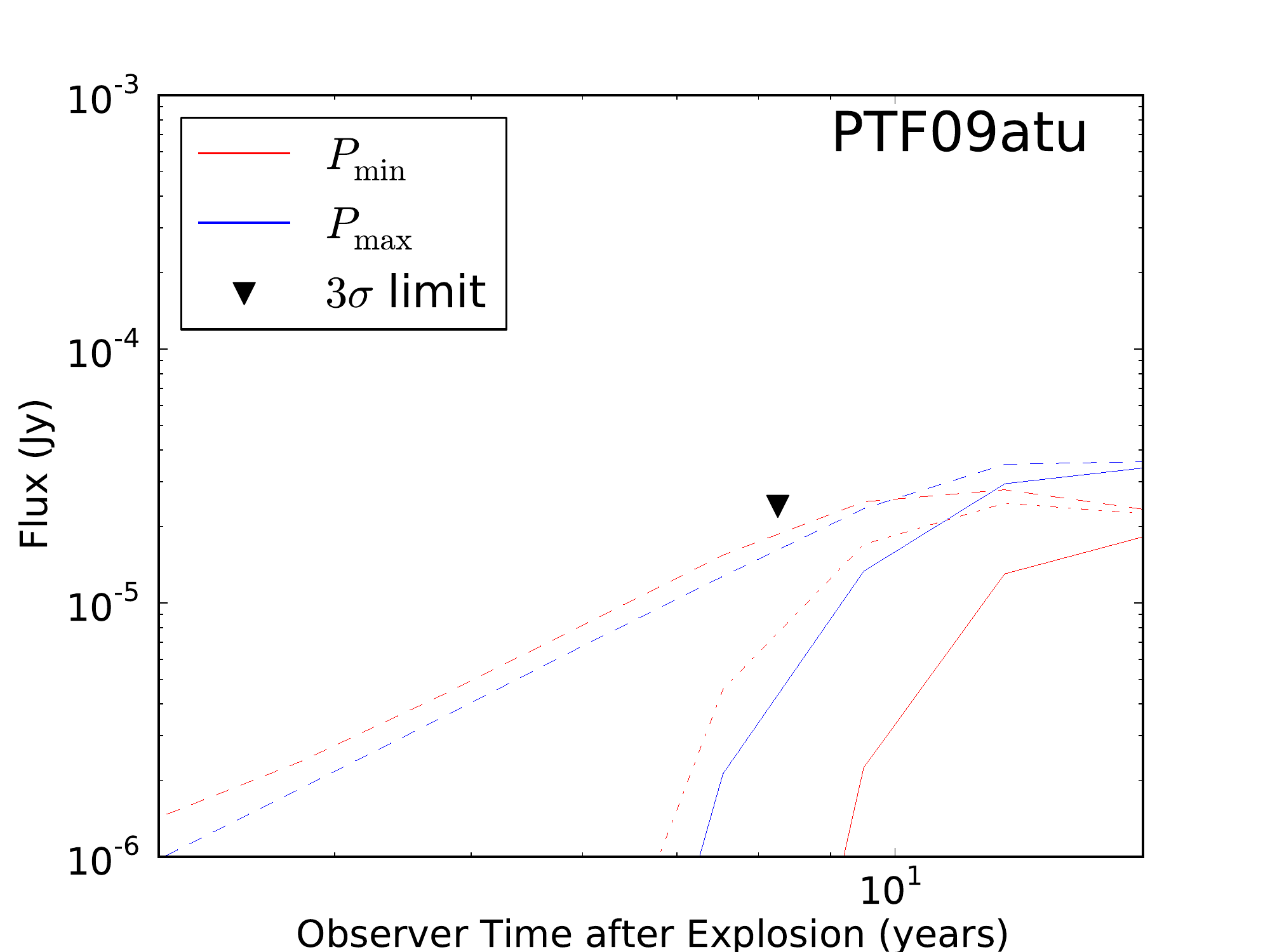}    
    \caption{Comparison of observed 3~GHz flux limits and expected flux from a range of pulsar-driven models constrained by the optical emission. Each panel shows a specific SLSNe-I and are ordered from left to right, starting at top: SN2005ap, SN2007bi, SN2006oz, PTF09cnd, SN2010kd, SN2010gx, PTF09cwl, SN2011ke, and PTF09atu. The black triangle show the 3$\sigma$\ flux limit, the red and blue lines show the models assuming $P_{\rm min}$ and $P_{\rm max}$ parameter sets, respectively, and the solid lines indicate the modeled radio flux assuming absorption, the dashed lines indicate the curve with no absorption, and the dash-dotted $P_{\rm min}$ line indicating a model with 40\% ionization.}
    \label{fig:lnu}%
\end{figure*}

\subsection{Non-detections}

Figure \ref{fig:lnu} shows the $3\sigma$\ upper limit on luminosity for the higher sensitivity observation for undetected SLSN shown in Table \ref{tab:obs} as a function of time since explosion. Table \ref{tbl:snparam} lists the magnetar and ejecta parameters that fit the optical light curves and are used for modelling radio emission.

\begin{table*}[tb]
\begin{center}
\begin{tabular}{|c|c|c|c|c|c|c|} \hline
Name &  $B_{13}$ at 1 ms & $M_{\rm ej}$ at 1 ms & $P_{\text{max}}$ & $B_{13}$ at $P_{\text{max}}$ & $M_{\rm ej}$ at $P_{\text{max}}$ & Data Reference \\ 
 & (G) & ($M_{\sun}$) & (ms) & (G) & ($M_{\sun}$) & \\ \hline
SN2005ap & 3.0 & 7 & 1.4 & 2.0 & 2.0 & \cite{2007ApJ...668L..99Q} \\
SN2007bi & 4.0 & 25 & 2.2 & 2.0 & 5.5 & \cite{2009Natur.462..624G} \\
SN2006oz & 5.0 & 12.5 & 2.0 & 2.0 & 2.5 & \cite{Lel} \\
PTF09cnd & 2.0 & 14 & 1.0 & 2.0 & 14 & \cite{2011Natur.474..487Q} \\
PTF10hgi & 14 & 15 & 4.2 & 4.0 & 2.0 & \cite{2013ApJ...770..128I} \\
SN2010kd & 4.7 & 25 & 2.4 & 2.0 & 4.0 & \cite{2012AAS...21943604V} \\
SN2010gx & 4.5 & 10.0 & 1.6 & 3.5 & 3.5 & \cite{2010ApJ...724L..16P} \\
PTF09cwl & 2.0 & 12 & 1.5 & 1.7 & 3.5 & \cite{sousa} \\ 
SN2011ke & 7.5 & 9.5 & 2.4 & 2.9 & 1.3 & \cite{2013ApJ...770..128I} \\
PTF09atu & 3.0 & 14 & 1.6 & 2.0 & 4.5 & \cite{2012PASP..124..668Y} \\\hline 
\end{tabular}
\caption{Model parameters from fits to the optical light curves for those SLSNe-I with sufficient optical data to constrain those models.  Periods were investigated from 1.0 ms to $P_{\text{max}}$, with any period above $P_{\text{max}}$ either not having enough luminosity, having too slow a decline, or having a shape inconsistent with the observed data. The uncertainty on these parameters is 5-10\% each.}
\label{tbl:snparam}
\end{center}
\end{table*}

The observational constraints on these models, which are summarized in Table \ref{tbl:modviable}, are as follows:

\begin{itemize}
    \item PTF09atu: We were not able to exclude any of the models, even those with no absorption. This is likely because PTF09atu is the furthest and youngest SLSNe-I in our sample.
    \item SN2007bi, PTF09cnd, SN2010kd, and PTF09cwl: Models with absorption are still viable to explain these SLSNe, but models with no absorption are excluded. The amount of absorption needed to be consistent with the model varies by supernova; PTF09cnd and PTF09cwl both need only a small amount of absorption to be consistent, while SN2007bi and SN2010kd both require more. All of them are consistent with a $P_{\rm min}$ model with 40\% ionized ejecta, like PTF10hgi.
    \item SN2010gx and SN2011ke: Both of these supernovae exclude models without absorption as well, and the $P_{\rm min}$ model would require a large amount of absorption in order to be consistent with observations, more than the best fit model for PTF10hgi.  The $P_{\rm max}$ model is also completely excluded for these two supernovae, so a faster spinning pulsar with larger magnetic field and ejecta mass is required to be consistent.
    \item SN2006oz: Free-free absorption here is predicted to be small, regardless of the pulsar parameters, mostly due to the age of the system. The $P_{\rm max}$ model is excluded by these observations, while the $P_{\rm min}$ model is still viable, even though the emission is predicted to be at or after the peak.  Only a small reduction in period from the $P_{\rm max}$ model would be required to make the model viable, however, since the predicted emission has almost the same flux as our 3$\sigma$ limit.
    \item SN2005ap: None of the models are consistent with our observations, as they all overpredict the expected emission. There are three likely reasons for this: this SLSN-I is not magnetar-driven; the electron injection spectrum is not broad and Crab-like \citep[e.g.][]{2010ApJ...715.1248T, 2013MNRAS.429.2945T}, but sharply peaked at higher energies (Omand et al. in prep); or the ejecta are more heavily ionized than predicted. \cite{2018MNRAS.tmp.2301M} predicts at most singly ionized species, but assumes 10 $M_\odot$ of ejecta. However, SN2005ap is best modeled with 2--7 $M_\odot$ of ejecta and \cite{2018ApJ...864L..36M} finds evidence for higher oxygen lines in SN2012au, a putative magnetar-driven supernova. Given these points, the ejecta may become more ionized on a timescale of $\sim$ 5 years. Free-free absorption outside the ejecta could also suppress the emission further.
\end{itemize}


\begin{table*}[tb]
\begin{adjustwidth}{-.5in}{-.5in} 
\begin{center}

\begin{tabular}{|c|c|c|c|c|c|} \hline
 Name &  $P_{\text{min}}$ abs & $P_{\text{min}}$ unabs &  $P_{\text{max}}$ abs & $P_{\text{max}}$ unabs & $P_{\text{min}}$ w/ PTF10hgi-like abs \\ \hline 
SN2005ap & Excluded & Excluded & Excluded & Excluded & Excluded \\
SN2007bi & Viable & Excluded & Viable & Excluded & Viable \\
SN2006oz & Viable & Viable & Excluded & Excluded & Viable \\
PTF09cnd & Viable & Excluded & Viable & Excluded & Viable \\
SN2010kd & Viable & Excluded & Viable & Excluded & Viable \\
SN2010gx & Viable & Excluded & Excluded & Excluded & Excluded \\
PTF09cwl & Viable & Excluded & Viable & Excluded & Viable \\
SN2011ke & Viable & Excluded & Excluded & Excluded & Excluded \\
PTF09atu & Viable & Viable & Viable & Viable & Viable \\\hline 
\end{tabular}
\end{center}
\end{adjustwidth}
\caption{A summary of viability of parameter sets for radio flux calculations. Sets of models were made for $P_{\text{min}}$ and $P_{\text{max}}$ and minimal and maximal free-free absorption opacity, as well as for PTF10hgi-like absorption.}
\label{tbl:modviable}
\end{table*}

\section{Conclusions}

We reported on new VLA observations to test the hypothesis that SLSNe-I are powered by young pulsars or magnetars. Of the ten SLSNe-I observed, we detect one, PTF10hgi, which supports earlier results and the argument it is a magnetar-powered supernova \citep{2019arXiv190110479E}. The detections of PTF10hgi are most consistent with the fastest-spinning magnetar model with minimal free-free absorption, for microphysical parameters similar to those of Galactic pulsar wind nebula \citep{2016MNRAS.461.1498M, 2018MNRAS.474..573O}.  The detection is also inconsistent with models with a low electron-injection Lorentz factor, which is typical for models of the persistent source of FRB 121102.  This may imply that these two sources have different electron acceleration mechanisms, or that the acceleration mechanism becomes less powerful over time, since the pulsar in FRB 121102 is expected to be older than that of PTF10hgi.

We measure upper limits for the radio luminosity of the other nine SLSNe-I. In general, these limits favour models with faster spins, higher magnetic fields, larger ejecta mass, and significant free-free absorption. This is in contrast to the best model for PTF10hgi.

While there may well be multiple mechanisms to power SLSNe-I, the young pulsar model predicts an increase in flux for all SLSNe-I in this sample \citep[][Omand et al. in prep]{2018MNRAS.474..573O}. Repeating these observations with the same sensitivity in 5--10 years would allow some constraint on pulsar parameters and six of them (SN2005ap, PTF09cnd, PTF09cwl, SN2010kd, SN2010gx, and SN2011ke) are predicted to be detectable under a range of scenarios. Observations today with more sensitive instruments (e.g. MeerKAT or SKA1) would also be likely to detect or better constrain the nature of the compact object.

PTF10hgi and the luminous radio source associated with FRB 121102 may be the first examples of $<100$~yr old pulsars \citep[c.f.,][]{2000ApJ...542L..37G, 2017JPhCS.932a2006D}. Radio observations of SN1986J also imply the existence of a compact object of yet unknown nature \citep{2017ApJ...839...10B, 2017ApJ...851....7B, 2017ApJ...851..124B}. Aside from their extreme luminosities, the former two sources are consistent with relatively flat radio spectra below 10 GHz. New observations of FRB 121102 would test whether the persistent radio source evolves in a similar manner as PTF10hgi. Similarly, broader spectral observations of PTF10hgi would test whether it has a similar spectral break as FRB 121102 \citep{2017Natur.541...58C}.

If other pulsar-powered supernovae can be identified, the radio properties can be used to study the birth properties of pulsars. The radio measurements of PTF10hgi sugggest it is powered by a pulsar born with a spin near the break-up period of 1~ms. Meanwhile, \citet{2017ApJ...839L...3K} used the properties of FRB 121102 to estimate a birth spin period of $\lesssim$ a few ms and age of 10-100 years. However, estimates of birth spin period are somewhat degenerate with magnetic field, photon absorption processes and more. Radio observations of known SLSN-I allows us to use the known age and optical light curve in modeling. Ultimately, we may be able to connect these young magnetars to the FRB phenomenon, which allows a host of new observational constraints, such as ejecta mass, age, and potentially spin period \citep{2016ApJ...824L..32P}.

\acknowledgments

We thank the VLA staff for their support of these challenging observations.
This research made use of Astropy,\footnote{http://www.astropy.org} a community-developed core Python package for Astronomy \citep{astropy:2013, astropy:2018}.

C.J.L.\ acknowledges support under NSF grant 1611606.
K. M.\ acknowledges financial support from the Alfred P. Sloan Foundation and NSF grant PHY-1620777.
K. K. acknowledges financial support from JSPS KAKENHI grant 18H04573 and 17K14248.
C. M. B. O. has been supported by the Grant-in-aid for the Japan Society for the Promotion of Science (18J21778).
S. B.-S, .K. A., and J. L.\ acknowledge support from NSF grant 1714897.
T. J. W. L. ackowledges research carried out at the Jet Propulsion Laboratory, California Institute of Technology, under a contract with the National Aeronautics and Space Administration. The NANOGrav project receives support from National Science Foundation (NSF) Physics Frontier Center award number 1430284.

\vspace{5mm}
\facilities{EVLA}

\software{rfpipe \citep{2017ascl.soft10002L}, astropy \citep{astropy:2013,astropy:2018}}

\bibliographystyle{aasjournal}

\end{document}